\documentclass[eqsecnum,nofootinbib,preprintnumbers]{revtex4}
\usepackage{bm,latexsym,amsmath,amssymb,amsfonts,mathrsfs,theorem,color,comment}
\usepackage[dvipdfmx]{hyperref}

\newcommand{\ma}[1]{\mbox{$\mathcal{#1}$}}

\newcommand{\qed}{\hfill $\Box$}
\newcommand{\D}{{\rm d}}

\newcommand{\we}{\wedge}

\def\Bi#1#2{\left(\begin{array}{c}#1\\#2\end{array}\right)}

{{\upshape}
\newtheorem{Prop}{Proposition}}
{\theorembodyfont{\upshape}
}
{\theorembodyfont{\upshape}
}

\begin{document}

\title{Lovelock black holes with non-constant curvature horizon}
\author{Seiju Ohashi${}^{1}$ and Masato Nozawa${}^{2}$}
\affiliation{${}^{1}$Theory Center, KEK, Tsukuba 305-0801, Japan}
\affiliation{${}^{2}$Dipartimento di Fisica, Universit\`a di Milano, and INFN, Sezione di Milano, Via G. Celoria 16, 20133 Milano, Italia}

\begin{abstract}
This paper studies a class of $D=n+2(\ge 6)$ dimensional solutions to Lovelock gravity that is described by the warped product of a two-dimensional Lorentzian metric and an $n$-dimensional Einstein space. Assuming that the angular part of the stress-energy tensor is proportional to the Einstein metric, it turns out that the Weyl curvature of an Einstein space must obey two kinds of algebraic conditions. We present some exact solutions satisfying these conditions. We further define the quasilocal mass corresponding to the Misner-Sharp mass in general relativity. It is found that the quasilocal mass is constructed out of the Kodama flux and satisfies the unified first law and the monotonicity property under the dominant energy condition. Making use of the quasilocal mass, we show Birkhoff's theorem and address various aspects of dynamical black holes  characterized by trapping horizons.  
\end{abstract}

\preprint{KEK-TH-1850, IFUM-1043-FT}
\maketitle

\section{Introduction}

In general relativity, or more general gravitational theories admitting diffeomorphism invariance, 
the role played by the mass is quite different from that in other branches of physics. 
The weak equivalence principle makes it  impossible to construct a well-defined ``local'' gravitational mass since it is always possible to set the local gravitational energy to vanish by working in a local Lorentz frame. This conceptual obstacle forced us to focus upon globally defined conserved quantities, e.g. the Arnowitt-Deser-Misner (ADM) mass in an isolated system~\cite{Arnowitt:1959ah}. However, one can circumvent this difficulty when the spacetime admits a spherical symmetry. In this case, the gravitational degrees of freedom is localized because of the absence of gravitational waves and it turns out to be extremely useful to define the ``quasilocal'' mass referring to the compact and orientable surfaces, as first demonstrated by Misner and Sharp~\cite{Misner:1964je}.  A number of local geometric properties of spacetime are encoded in the Misner-Sharp mass~\cite{Hayward:1994bu}, including the causal property of central singularities, trapped surfaces and asymptotic charges. It follows that the Misner-Sharp mass 
provides a fruitful venue for dexterously capturing the dynamical aspects of gravitational collapse.

Inspired by recent advances of string theory, many people have tried to extend Einstein's gravity. 
A covariant gravitational theory constructed by Lovelock~\cite{Lovelock}  is a natural  extension of general relativity into $D(\ge 5)$ dimensions.  The most appealing and characteristic feature of Lovelock gravity inherited from general relativity is that the field equations continue to remain second order, irrespective of the fact that they are accompanied by higher-order polynomials of curvature tensors. This traces back to the topological interpretation of each Lovelock term as the dimensionally continued Euler densities, 
allowing the understanding of Lovelock terms in the context of the BRST cohomology~\cite{Cnockaert:2005jw}. 
Therefore, there appear no ghost degrees of freedom at the linearized level ~\cite{Zwiebach:1985uq,Zumino:1985dp}, and the Lovelock gravity is a classically well-posed gravitational theory (see  \cite{Izumi:2014loa,Reall:2014sla} for a recent discussion opposing this belief). Apart from this theoretical aesthetic beauty, the quadratic Lovelock term dubbed the ``Gauss-Bonnet'' term arises as a low-energy effective action in heterotic string theory~\cite{Gross,Gross2,Metsaev}. This motivates us to explore quantum aspects of higher-curvature terms as well in light of AdS/CFT correspondence~\cite{Brigante:2008gz}.

Since higher-curvature terms come into play where the gravitational force becomes very strong, 
black holes are the best test beds in which deviations from general relativity are capitally encoded. 
The complexity of Lovelock field equations has restricted the analysis especially to the spacetimes with  a high degree of symmetry. Among other things, many works have focused upon the spacetime which is 
the warped product of two-dimensional Lorentzian spacetime and an $n$-dimensional maximally symmetric space. The simplest solution is the spherically symmetric black hole found by Whitt~\cite{Whitt:1988ax},  
as a generalization of the Schwarzschild solution in general relativity. Thermodynamics~\cite{Myers:1988ze,Cai:2003kt} and gravitational instabilities~\cite{Takahashi:2009dz,Takahashi:2009xh,Takahashi:2010ye,Takahashi:2010gz,Takahashi:2011qda,Takahashi:2012np} of this type of black hole have been intensively studied.  This class of metrics also contains the Tolman-Bondi inhomogeneous dust spacetime \cite{maeda2006b,Ohashi:2011zza,Ohashi:2012wfa} and the Vaidya-type radiating solution~\cite{Kobayashi:2005ch,Maeda:2005ci,Nozawa:2005uy,Cai:2008mh}, both of which describe the gravitational collapse. The examinations of gravitational collapse have revealed that the global structure turns out to be quite different from that encountered in general relativity, and a peculiar type of massive singularity emerges in every odd dimension. In these analyses, the generalized Misner-Sharp quasilocal mass~\cite{Maeda:2007uu,Maeda:2011ii} plays an essential role as in general relativity. In this sense, the  Misner-Sharp quasilocal mass is more advantageous than the Brown-York  quasilocal mass~\cite{by1993} constructed based upon the Hamiltonian formalism.

In general relativity, the $n$-dimensional maximally symmetric space can be replaced by arbitrary Einstein spaces, since their Weyl curvature fails to contribute to Einstein's equations.\footnote{The replacement to the Einstein space has a significant impact upon the linear instability of black holes~\cite{Gibbons:2002pq}.} In Lovelock gravity, on the other hand, the Weyl tensor appears explicitly in field equations and, therefore, the generic Einstein manifold fails to satisfy the vacuum field equations of the warped metric. The condition that this type of metric admit vacuum solutions imposes two conditions upon the Weyl curvature of the Einstein space. When one takes the Einstein space satisfying these conditions, the causal structures of the black hole considerably differ from those  with maximally symmetric horizons, as argued in the Gauss-Bonnet gravity \cite{Dotti:2005rc,Maeda:2010bu,Pons:2014oya} and in the third-order Lovelock gravity \cite{Farhangkhah:2014zka}. This motivates our present attempt to explore the conditions for an Einstein horizon in general Lovelock gravity, by extending the analysis in \cite{Dotti:2005rc,Maeda:2010bu,Pons:2014oya,Farhangkhah:2014zka}\footnote{The solutions to special cases of Lovelock gravity with a more general base space were also studied in \cite{Dadhich:2015nua,Anabalon:2011bw,Dotti:2010bw,Oliva:2012ff}.}.  

In this paper, we generalize the previous studies \cite{Dotti:2005rc,Maeda:2010bu,Pons:2014oya,Farhangkhah:2014zka} into Lovelock gravity where the spacetime consists of the warped product of the two-dimensional Lorentzian metric and the $n$-dimensional Einstein space. We find that the Weyl tensor of $n$-dimensional Einstein space must obey two conditions. We find that a dozen new Einstein spaces turn out to satisfy the Lovelock field equations.
We extend the definition of the Misner-Sharp type quasilocal mass adapted to the present context. It turns out that the quasilocal mass displays some desirable physical properties under suitable energy conditions, provided that some conditions on the Weyl curvature and the coupling coefficients of Lovelock action are satisfied. Using the quasilocal mass, we further explore the properties of trapping horizons and their thermodynamics.

The present paper proceeds as follows. In the next section, we give a brief review of Lovelock gravity and derive field equations under the setup described above. In Sec.~\ref{sec:QLM}, we define a quasilocal mass as a generalization of the Misner-Sharp mass. We explore a number of properties of dynamical black holes defined by the trapping horizons using the quasilocal mass in Sec.~\ref{sec:trapping}. Final remarks are described in Sec.~\ref{sec:conclusion}. 
In the Appendix, we give a variety of exact solutions for vacuum and electrovacuum cases.  We follow the conventions of Wald's textbook~\cite{Wald} for curvature tensors.

\section{SetUp}
\label{sec:setup}

The action of Lovelock gravity in $D$ dimensions is~\cite{Lovelock} 
\begin{align}
\label{action}
S=\frac{1}2\int \D ^D x \sqrt{-g} \left(
\sum_{m=1}^k\frac{1}{2^m}\frac{a_m}m \delta ^{\mu_1\mu_2 \cdots \mu_{2m-1}\mu_{2m}}
_{\nu_1\nu_2 \cdots \nu_{2m-1}\nu_{2m}} R_{\mu_1\mu_2 }{}^{\nu_1\nu_2} \cdots 
R_{\mu_{2m-1}\mu_{2m} }{}^{\nu_{2m-1}\nu_{2m}} 
+2\Lambda
\right)+S_{\rm mat}\,, 
\end{align}
where $a_m$ are real constants  and we set $a_1=1$ and $8\pi G=1$. 
$k $ is given by $k\equiv \lfloor(D-1)/2\rfloor$, where the symbol $\lfloor x\rfloor$ denotes the integer part of $x$.\footnote{
Note that some literatures have employed the convention $k=\lfloor D/2\rfloor$ different from ours.   
Since the $D/2$th term in even dimensions amounts to the topological invariant, it fails to contribute to the field equation. Hence, both of these conventions do not make any physical difference. 
} 
$S_{\rm mat}$ is the action for the matter field, 
$\Lambda $ is a  cosmological constant and $\delta$ denotes the totally antisymmetric 
product of Kronecker delta normalized by 
\begin{align}
\label{}
\delta ^{\mu_1\mu_2 \cdots \mu_{m-1}\mu_{m}}
_{\nu_1\nu_2 \cdots \nu_{m-1}\nu_{m}}= 
\delta ^{\mu_1}_{\nu_1}\delta ^{\mu_2}_{\nu_2}
 \cdots \delta^{\mu_{m-1}}_{\nu_{m-1}}\delta^{\mu_m}_{\nu_m}+{\rm cyclic} \,.
\end{align}
The gravitational field equations derived from the action (\ref{action}) reads 
\begin{align}
\label{EOM}
\ma G_{\mu\nu}= T_{\mu\nu} \,,
\end{align}
where $T_{\mu\nu}=-2 \delta S_{\rm mat}/\delta g^{\mu\nu}$ 
describes the stress tensor of the matter fields. 
The Lovelock tensor $\ma G_{\mu\nu}$ is given by
\begin{align}
\label{}
\ma G^\mu{}_\nu = -\sum_{m=1}^k \frac 1{2^{m+1}}\frac{a_m}{m}
\delta ^{\mu \rho_1\rho_2 \cdots \rho_{2m-1}\rho_{2m}}_{
\nu \sigma_1\sigma_2 \cdots \sigma_{2m-1}\sigma_{2m}}
R_{\rho_1\rho_2}{}^{\sigma_1 \sigma_2}\cdots 
R_{\rho_{2m-1}\rho_{2m}}{}^{\sigma_{2m-1} \sigma_{2m}}\,,
\end{align}
obeying the Bianchi identity $\nabla^\nu \ma G_{\mu\nu}=0$. 
A notable feature of Lovelock gravity is that the equations of motion involve 
no more than the third derivative of the metric.

In this paper, we consider the $D=n+2$-dimensional spacetime $(\mathcal{M}^D,g_{\mu \nu})$ for which the metric takes the cross product of the two-dimensional orbit spacetime $(M^2,g_{ab})$ and the $n$-dimensional Einstein space $(\mathcal{K}^n,\gamma_{ij})$. Namely, the local metric reads  
\begin{align}
\label{metric}
\D s^2=g_{ab}(y)\D y^a\D y^b+r^2(y)\gamma_{ij}(x^k)\D x^i\D x^j \,, 
\end{align}
where $r$ is the scalar on $M^2 $ corresponding to the warp factor. 
Indices $a, b,...$ run over $0,1$ and $i,j,...$ correspond to those of the Einstein space. 
The Ricci tensor of Einstein space $(\mathcal{K}^n,\gamma_{ij})$ reads $R_{ij}[\gamma] =(n-1) \kappa \gamma_{ij}$, where $\kappa $ is the constant normalized by $\kappa=\pm 1, 0$. 
We assume that $(\mathcal{K}^n,\gamma_{ij})$ is compact with the area $V_n^\kappa$, and that 
$(M^2, g_{ab})$ is a time-orientable Lorentzian manifold. 
The Riemann tensor of (\ref{metric}) decomposes into 
\begin{align}
\label{}
R_{abcd}={}^{(2)}R_{abcd}\,, \qquad R_{aibj}=-r(D_aD_b r) \gamma_{ij} \,, \qquad 
R_{ijkl} =r^2 [R_{ijkl}[\gamma] -2(Dr)^2\gamma_{i[k}\gamma_{l]j} ] \,, 
\end{align}
where $D_a$ is a covariant derivative with respect to $g_{ab}$ and $(Dr)^2 \equiv g^{ab}(D_a r)(D_br)$. 
The suffix ``2'' is attached with quantities of $M^2$, which are distinguished by those of 
$(\ma K^n, \gamma_{ij})$ represented by $[\gamma]$. 
One can express the Weyl tensor of the Einstein space as 
\begin{align}
\label{}
R_{ij}{}^{kl} [\gamma] =C_{ij}{}^{kl}[\gamma] +\kappa \delta^{kl}_{ij} \,. 
\end{align}
When ($\ma K^n, \gamma_{ij}$) is maximally symmetric, 
the present setup collapses to the case analyzed in \cite{Maeda:2011ii}. 
Note that the Einstein space is necessarily maximally symmetric in $n=3$, 
the nontriviality arises in $D\ge 6$ dimensions.

\subsection{Field equations}

In the following calculation, we frequently use the quantities $W(s)^{i}{}_j$ and $W(s)$, which are defined as
\begin{equation}
W(s)^{i}{}_j\equiv 
\begin{cases}
\delta^i_j & s=0\\
\delta^{i i_1 i_2 \dots i_{2s-1}i_{2s}}_{j{j_1 j_2 \dots j_{2s-1}j_{2s}}}C_{i_1 i_2}{}^{j_1 j_2}[\gamma]\dots C_{i_{2s-1}i_{2s}} {}^{j_{2s-1}j_{2s}}[\gamma] & s\geq 1\,,
\end{cases}
\end{equation}
and 
\begin{equation}
W(s)\equiv 
\begin{cases}
1 & s=0\\
\delta^{ i_1 i_2 \dots i_{2s-1}i_{2s}}_{{j_1 j_2 \dots j_{2s-1}j_{2s}}}C_{i_1 i_2}{}^{j_1 j_2}[\gamma]\dots C_{i_{2s-1}i_{2s}}{}^{j_{2s-1}j_{2s}}[\gamma] & s\geq 1.
\end{cases}
\end{equation}
$W(s)_{ij}$ are symmetric tensors on ($\ma K^n, \gamma_{ij}$). We also define  
\begin{align}
\left( \begin{matrix} m \\ l\end{matrix}\right)  \equiv {}_mC_{l} .
\end{align}
By a straightforward computation, 
the Lovelock tensor decomposes into
\begin{align}
\label{Gab}
\ma G^a{}_b=& \sum_{m=1}^k\sum_{l=0}^{m-1}
\frac{a_m 2^{l-m+1}}{r^{2m-2}} \Bi{m-1}{l}  \left[
\delta^a{}_b \frac{D^2 r}r -\frac{D^a D_b r}r -\delta ^a{}_b (n-2m+1)\frac{\kappa-(Dr)^2}{2(l+1)r^2}
\right]
\nonumber \\
& \times \left(\prod_{p=0}^{2l}(n-2m+2+p)\right)(\kappa-(Dr)^2) ^l W(m-1-l) \nonumber \\
&-\sum_{m=1}^k \frac{1}{2^{m+1}}\frac{a_m}m \delta^a{}_b \frac{W(m)}{r^{2m}} 
+\Lambda \delta ^a{}_b \,, \\
\label{Gij}
\ma G^i{}_j=& \sum_{m=1}^k \frac{a_m}{2^{m-1}}\frac{D^2 r}{r^{2m-1}}\left[
\sum_{l=0}^{m-1} 2^l \Bi{m-1}{l}\left(\prod_{p=0}^{2l }(n-2m+1+p)\right)(\kappa-(Dr)^2)^l W(m-1-l)^i{}_j
\right]\nonumber \\
&-\sum_{m=1}^k \frac{a_m}{2^m}\frac{{}^{(2)} R}{r^{2(m-1)}} \left[
\sum_{l=0}^{m-1}\frac{2^l }{n-(2m-1)} \Bi{m-1}{l}\left(\prod_{p=0}^{2l}
(n-(2m-1)+p)\right)(\kappa-(Dr)^2)^lW(m-1-l) ^i{}_j
\right] \nonumber \\
&-\sum_{m=1}^k\frac{(m-1)a_m}{2^{m-2}}\frac{\delta^{ab}_{cd}(D_aD^cr)( D_b D^d r)}{r^{2(m-1)}}\left[(n-2m+2)\sum_{l=0}^{m-2}2^l \Bi{m-2}{l}
\nonumber \right.\\ & \qquad \times  \left.
\left(\prod_{p=0}^{2l}(n-(2m-3)+p)\right) (\kappa-(Dr)^2)^lW({m-2-l})^i{}_j\right]
\nonumber \\
&-\sum_{m=1}^k \frac{a_m}{2^{m+1}m r^{2m}} \left[\sum_{l=0}^m \frac{2^l }{n-2m} 
\Bi{m}{l}
\left(\prod_{p=0}^{2l}(n-2m+p)\right)(\kappa-(Dr)^2)^l W(m-l)^i{}_j\right]
+\Lambda \delta^i{}_j \,. 
\end{align}

For a generic Einstein space $(\ma K^n, \gamma_{ij})$, 
$W(m)$ are functions dependent on the coordinates $x^i $ and the 
(trace-free part of) symmetric tensor $W(m)_{ij}$ is nontrivial. 
In that case, the Lovelock tensor $\ma G^\mu {}_\nu$ involves the convoluted 
coordinate dependence on $x^i$, as well as the dependence on $y^a$.   
In order to avoid these technical difficulties and make the discussion focused, 
we impose in this paper the following two conditions on  $(\ma K^n, \gamma_{ij})$:
\begin{subequations}
\label{Wcond}
\begin{align}
W(m)^i{}_j&=\frac{n-2m}n \delta^i{}_j W(m) \,, \label{Wcond1}\\
 W(m)&={\rm const. } \label{Wcond2}
\end{align}
\end{subequations}
With these conditions, the $x^i$ dependence of $\ma G^\mu{}_\nu$ drops out except for 
the contribution stemming from the metric $\gamma_{ij}$. In \cite{Dotti:2005rc} a similar condition was imposed in the $m=2$ case. 
On account of the dimensionally dependent Lovelock identities~\cite{Lovelock2}, 
the constraint (\ref{Wcond1}) is automatically satisfied for $m\ge \lfloor (n+1)/2\rfloor=k$. 
Obviously, the conditions (\ref{Wcond}) restrict the permissible horizon topologies for static black holes.  Appendix~\ref{app:ex} illustrates some explicit examples of Einstein spaces satisfying 
(\ref{Wcond}). 

The stress tensor compatible with these assumptions, therefore, reads 
\begin{align}
\label{SETensor}
T_{\mu\nu} \D x^\mu \D x^\nu = (\hat T_{ab}(y) -P(y) g_{ab}) \D y^a \D y^b 
+r^2 (y) p(y) \gamma_{ij} (x)\D x^i \D x^j \,, 
\end{align}
where 
\begin{align}
\label{PhatT}
P\equiv -\frac 12 T^a{}_a \,, \qquad 
g^{ab} \hat T_{ab}=0 \,. 
\end{align}
The stress-energy tensor obeys the 
conservation law
\begin{align}
\label{SEcons}
0=\frac 1{r^n}D_b [r^n (\hat T^{ab}-P g^{ab})] -\frac{n}{r} (D^a r) p \,.
\end{align}

Under these settings, the Lovelock field equations are given by
\begin{align}
\label{Loveeq_tf}
\hat T_{ab}=&-\sum_{m=1}^k\sum_{l=0}^{m-1}
\frac{a_m 2^{l-m+1}}{r^{2m-1}} \Bi{m-1}{l} \left(\prod_{p=0}^{2l}(n-2m+2+p)\right)(\kappa-(Dr)^2) ^l W(m-1-l)
\nonumber \\ & \times
\left(
D_a D_b r-\frac 12 D^2 r g_{ab} 
\right) \,, \\ 
\label{Loveeq_tr}
P=&- \sum_{m=1}^k\sum_{l=0}^{m-1}
\frac{a_m 2^{l-m}}{r^{2m-2}} \Bi{m-1}{l}  \left[
\frac{D^2 r}r -(n-2m+1)\frac{\kappa-(Dr)^2}{(l+1)r^2}
\right]
\nonumber \\
& \times \left(\prod_{p=0}^{2l}(n-2m+2+p)\right)(\kappa-(Dr)^2) ^l W(m-1-l) 
+\sum_{m=1}^k \frac{1}{2^{m+1}}\frac{a_mW(m)}{m r^{2m}} 
-\Lambda  \,, \\
\label{Loveeq_ij}
p=& \sum_{m=1}^k \frac{a_m}{2^mnr^{2(m-1)}}\left[{2}\frac{D^2 r}{r}-\frac{{}^{(2)}R}{n-(2m-1)}\right]
\sum_{l=0}^{m-1} 2^l \Bi{m-1}{l}\left(\prod_{p=0}^{2l +1}(n-2m+1+p)\right)
(\kappa-(Dr)^2)^l W(m-1-l)
\nonumber \\
&-\sum_{m=1}^k\frac{(m-1)a_m}{2^{m-2}n}\frac{\delta^{ab}_{cd}(D_aD^cr)( D_b D^d r)}{r^{2(m-1)}}\sum_{l=0}^{m-2}2^l \Bi{m-2}{l}
\left(\prod_{p=0}^{{2l+2}}(n-2m+2+p)\right) (\kappa-(Dr)^2)^lW({m-2-l})
\nonumber \\
&-\sum_{m=1}^k \frac{a_m}{2^{m+1}m r^{2m}} \left[\sum_{l=0}^m \frac{2^l(n-2m+2l) }{n(n-2m)} 
\Bi{m}{l}
\left(\prod_{p=0}^{2l}(n-2m+p)\right)(\kappa-(Dr)^2)^l W(m-l)\right]
+\Lambda\,. 
\end{align}
If $r$ is not constant, the angular part of field equation (\ref{Loveeq_ij}) follows from 
(\ref{Loveeq_tf}), (\ref{Loveeq_tr}) and  (\ref{SEcons}). 

In the above we assumed (\ref{Wcond}) to simplify the system. As far as the vacuum solution is concerned, the condition (\ref{Wcond})  actually follows from the consistency with the field equations, 
as shown in \cite{Dotti:2005rc} for Gauss-Bonnet gravity.

In the following sections, we shall discuss the general properties of the metric under suitable energy conditions,  without resorting to the exact solutions. 
In Appendix \ref{app:matter}, we give some exact solutions of physical interest.

\section{Quasilocal mass}
\label{sec:QLM}

In general relativity, the spacetime admitting spherical symmetry allows no freedom of gravitational radiations. This fact enables us to localize the gravitational energy and one can define the quasilocal mass~\cite{Misner:1964je} that plays an important role in the analysis of dynamics~\cite{Hayward:1994bu}.
 When $(\ma K^n , \gamma_{ij})$ is a maximally symmetric space, analogous quasilocal quantities have been generalized to Gauss-Bonnet~\cite{Maeda:2007uu} and to Lovelock gravities \cite{Maeda:2011ii}.
These definitions have been further extended to the case of Einstein spaces in the Gauss-Bonnet gravity \cite{Maeda:2010bu} and in the third-order Lovelock gravity~\cite{Farhangkhah:2014zka}.  Here we complete the series of research by studying the nonconstant curvature case in the full Lovelock gravity, which encompasses all the previous studies. 

By mimicking the quantity in Maeda's paper~\cite{Maeda:2010bu}, 
our proposed definition of  the quasilocal mass reads
\begin{align}
M (y)\equiv V^\kappa_{n}&\Bigg[ \sum_{m=1}^k\frac{1}{2^{m+1}}\frac{a_m}{m}\frac{r^{n-2m+1}}{(n-2m+1)}W(m)-\frac{r^{n+1}}{(n+1)}\Lambda \notag \\
           &+\sum_{m=1}^k  \sum_{l =0}^{m-1}a_m2^{l-m}\frac{r^{n-2m+1}}{l +1}\left( \begin{matrix} m-1 \\ l \end{matrix}\right) \left(\prod_{p=0}^{2l }(n-(2m-2)+p)\right)(\kappa -(Dr)^2)^{l +1}W(m-1-l) \Bigg] \,.
\label{MSmass}
\end{align}
Due to the assumption (\ref{Wcond2}), one may view the quasilocal mass as a scalar on 
($M^2, g_{ab}$). It is  constructed out of the areal radius $r$ and its first derivative, as well as the 
Weyl tensor of Einstein space. When the space ($\ma K^n , \gamma_{ij}$) is maximally symmetric, the above definition reduces to the one given in \cite{Maeda:2011ii}. It also recovers the well-defined Misner-Sharp mass~\cite{Misner:1964je} in general relativity ($a_{m\ge 2}=0$), and its generalization in Gauss-Bonnet gravity ($a_{m\ge 3}=0$)
~\cite{Maeda:2007uu,Maeda:2010bu}. 
Using (\ref{Gab}), one can easily verify that  $M$ satisfies the variation formula 
\begin{align}
\label{variation}
D_aM =V^\kappa_{n}r^{n}( \mathcal{G}_{ab}D^br -\mathcal{G}^b{}_bD_ar )\,.
\end{align}
This formula takes exactly the same form as those analyzed in previous studies~\cite{Maeda:2007uu,Maeda:2011ii,Maeda:2010bu}. This relation is of crucial importance in the following discussion.

\subsection{Locally conserved currents}

The physical meaning of $M$ is less clear in the geometric definition (\ref{MSmass}).  
In this section, we shall demonstrate that $M$ can be rebuilt in terms of a locally conserved energy flux. 

To proceed, let us first define the Kodama vector~\cite{Kodama:1979vn}
\begin{align}
\label{Kodama}
K^\mu \equiv -\epsilon^{\mu\nu}\nabla_\nu r \, ,
\end{align}
where $\epsilon_{\mu\nu}=\epsilon_{ab}(\D y^a)_\mu (\D y^b)_\nu $ and 
$\epsilon_{ab}$ is a volume element of ($M^2, g_{ab}$). 
This current can be viewed as a vector field on $M^2$ since $K^i =0$. 
The Kodama vector fulfills the following crucial property:
\begin{align}
\label{Knorm}
K^\mu K_\mu =-(\nabla r)^2 \,. 
\end{align}
This means that the Kodama vector is timelike (spacelike) in the untrapped (trapped) region and specifies the preferred time direction. 
Using the Kodama vector, one can also define a  Kodama current,
\begin{align}
\label{}
J^\mu \equiv -\ma G^\mu{}_\nu K^\nu \,, 
\end{align}
which is again a vector field on $M^2$. 
Using the Lovelock field equation (\ref{EOM}), one obtains $J^\mu =-T^\mu{}_\nu K^\nu$. Hence $J^\mu$ describes an energy flux. Because of the properties $K^a D_a r=0$ and $\ma G_{ab}D^a K^b=0$, 
one sees that these vectors are divergence free
\begin{align}
\label{divKJ}
\nabla_\mu K^\mu =0 \,, \qquad \nabla_\mu J^\mu =0 \,. 
\end{align}
Using $D_a r=- \epsilon_{ab}K^b$, 
one can easily derive the following relations, 
\begin{align}
\label{}
K^a =- r^{-n}\epsilon^{ab}D_b (V/V_n^\kappa) \,, \qquad 
J^a = -r^{-n}\epsilon^{ab}D_b( M/V_n^\kappa) \,,
\end{align}
where 
\begin{align}
\label{}
V\equiv & \frac{V^\kappa_n}{n+1}r^{n+1} 
\end{align}
is a weighted volume of $\ma K^n$. 
It follows that vector quantities $K^\mu $ and $J^\mu$ are the Hamiltonian vector fields with the corresponding Hamiltonian
$V$ and $M$. This expression makes the divergence-free property (\ref{divKJ}) rather manifest. 
It is also obvious to see
\begin{align}
\label{volume}
V=- \int _\Sigma K^\mu u_\mu \D \Sigma \,, 
\qquad 
M=- \int _\Sigma J^\mu u_\mu \D \Sigma \,, 
\end{align}
where $\Sigma $ is a $(D-1)$-dimensional hypersurface without an interior boundary and $u_\mu$ is a  future pointing unit normal to $\Sigma$ . 
This accomplishes our first aim to prove that $M$ is a quasilocal quantity associated with the locally conserved energy flux. The definition of  the Misner-Sharp quasilocal mass based upon the energy flux illustrates the direct physical relevance rather than the original geometric definition~(\ref{MSmass}).

\subsection{Unified first law}

The first law of thermodynamics is one of the fundamental laws in nature. Hence, 
the validity of the first law deserves a nice criterion for the well-defined mass. 
We can easily check from (\ref{variation}) that the following unified first law ~\cite{hayward1998} holds
\begin{align}
\label{U1st}
\D M =A\psi_a\D x^a+P\D V\,, 
\end{align}
where $P$ has been defined in (\ref{PhatT}), and 
\begin{align}
\label{psiA}
\psi^a \equiv &\, \hat T^a{} _bD^br\,, \qquad 
A \equiv  V^\kappa _nr^n \,. 
\end{align}
$A $ is the weighted area of Einstein space and is related to the volume (\ref{volume}) as $D_a V=A D_a r$.  As it turns out from the analysis of the next subsection, $\psi^a$ describes a momentum flux.   Therefore, (\ref{U1st}) represents the physical circumstance that the energy balance is compensated by the work term $P\D V$ and the energy inflow provided by $\psi^a$. The unified 1st law allows us to interpret $M$ as an energy contained in the closed surface enclosed by the geometric radius $r $.

\subsection{Birkhoff's theorem} 

The theorem of Birkhoff plays a significant role when one analyzes the gravitational collapse of a spherical body in general relativity. One of the characteristic features of Lovelock gravity is that Birkhoff's theorem (and modified versions thereof) continues to hold, as discussed in \cite{Maeda:2007uu,Maeda:2011ii,Zegers:2005vx,Deser:2005gr,Ray:2015ava}.

Let us consider the matter fields with $\hat T_{ab}=0$. 
Suppose that the first line of equation (\ref{Loveeq_tf}) is nonvanishing.\footnote{We shall not discuss in this paper the case where the first line of (\ref{Loveeq_tf}) identically vanishes. For these artfully chosen values of $a_m$ with a given Einstein space, there appears a solution for which the metric $g_{ab}$ is undetermined. For this class of metrics, we refer the reader to Refs.~\cite{Maeda:2007uu,Maeda:2011ii,Zegers:2005vx}.} This implies that 
\begin{align}
\label{CKV}
0=
D_a D_b r-\frac 12 D^2 r g_{ab}  \,. 
\end{align}
Assume that $D_a r$ does not vanish. Then, $D_a r$ describes a conformal Killing field on $M^2$, which implies 
\begin{align}
\label{}
D_a K_b =\frac 12 D^2 r \epsilon_{ab} \,. 
\end{align}
It follows that  $K^a $ is a Killing vector on $M^2$. Thanks to the property 
$\nabla_\mu K_\nu=D_a K_b(\D y^a)_\mu (\D y^b)_\nu$, 
$K^\mu$ describes a hypersurface-orthogonal Killing vector on $\ma M^D$,
\begin{align}
\label{}
K_{[\mu }\nabla_\nu K_{\rho]} =0 \,, \qquad \nabla_{(\mu }K_{\nu)}=0 \,. 
\end{align}
On account of the property (\ref{Knorm}), this means that the spacetime is static in the untrapped 
region. Hence, $\hat T_{ab}=0$ provides a sufficient condition for the validity of staticity (in the untrapped region).  This also justifies the physical interpretation of $\psi^a$ as a flux current, since it vanishes in static spacetimes.

In order to obtain the metric explicitly, let us concentrate on the vacuum spacetime ($P=p=0$) in what follows. 
Then, 
the variation formula (\ref{variation}) [or the unified first law (\ref{U1st})] immediately gives $M=\mu={\rm const}$.  
Let us introduce the coordinate $t$ by $K=\partial/\partial t$ and denote the norm of $K^\mu$ by 
$-f$, i.e, $K_\mu =-f \nabla_\mu t $.  Under the condition $(Dr)^2\ne 0$ one can use $r $ as the coordinate on $M^2$ conjugate to $t$, and 
the general solution reads
\begin{align}
\label{staticsol}
\D s_2^2 = -f(r)\D t^2+\frac{\D r^2}{f(r)} \,, 
\end{align}
where $f(r)$ satisfies 
\begin{align}
\label{Mass_static}
\mu = &V_n^\kappa \Biggl[\sum_{m=1}^k \frac{1}{2^{m+1}}\frac{a_m}m \frac{r^{n-2m+1}}{n-2m+1}
W(m)-\frac{r^{n+1}}{n+1}\Lambda \nonumber \\
& +\sum_{m=1}^k \sum_{l=0}^{m-1}2^{l-m}a_m \frac{r^{n-2m+1}}{l+1}\Bi{m-1}{l}
\left(\prod_{p=0}^{2l}(n-(2m-2)+p)\right) (\kappa-f(r))^{l+1}W(m-1-l)\Biggl]\,.
\end{align}
This is a simple Lovelock generalization of the Dotti-Gleiser solution~\cite{Dotti:2005rc} for 
Einstein-Gauss-Bonnet gravity. 

The case $r=r_0={\rm const.}$ also solves (\ref{CKV}). In this case, 
(\ref{Loveeq_ij}) gives that ${}^{(2)}R $ is constant, thus ($M^2 ,g_{ab}$) is 
a spacetime of constant curvature. 
The metric can therefore be written as 
\begin{align}
\label{Nariai}
\D s^2 =- (1 -\lambda x^2) \D t^2 +\frac{\D x^2}{1-\lambda x^2} 
+r_0^2 \gamma_{ij }\D x^i \D x^j \,,
\end{align}
where  $r_0$ and $\lambda$ satisfy the following relations
\begin{align}
\label{}
\Lambda=& \sum_{m=1}^k\sum_{l=0}^{m-1}
\frac{a_m 2^{l-m}}{r_0^{2m}} \Bi{m-1}{l}  \frac{n-2m+1}{l+1}
\kappa^{l+1} 
 \left(\prod_{p=0}^{2l}(n-2m+2+p)\right)W(m-1-l) 
\nonumber \\&
+\sum_{m=1}^k \frac{1}{2^{m+1}}\frac{a_mW(m)}{m r_0^{2m}} 
  \,, \\
\Lambda=& \sum_{m=1}^k \frac{a_m }{2^{m-1}nr_0^{2(m-1)}}
\frac{\lambda}{n-(2m-1)}
\sum_{l=0}^{m-1} 2^l \Bi{m-1}{l}\left(\prod_{p=0}^{2l +1}(n-2m+1+p)\right)
\kappa^l  W(m-1-l)
\nonumber \\
&+\sum_{m=1}^k \frac{a_m}{2^{m+1}m r_0^{2m}}\sum_{l=0}^m \frac{2^l(n-2m+2l) }{n(n-2m)} 
\Bi{m}{l}
\left(\prod_{p=0}^{2l}(n-2m+p)\right)\kappa^l W(m-l)\,. 
\end{align}
(\ref{Nariai}) describes a Nariai-type metric ${\rm (A)dS}_2\times \ma K^n$. 

\subsection{Physical properties of quasilocal mass}

In order to analyze the dynamics of the warped spacetime (\ref{metric}), it 
is advantageous to work in the double null coordinates 
\begin{align}
\D s^2=-2e^{-f(u,v)}\D u\D v+r^2(u,v)\gamma_{ij}\D x^i\D x^j\,,
\end{align}
where the orientation is fixed to be $\epsilon_{uv}>0$. 
Then we can see that the variation formula (\ref{variation}) may be cast into 
\begin{subequations}
\label{variation_uv}
\begin{align}
\partial_u M&=\frac{1}{n}V^\kappa_ne^fr^{n+1}\left( T_{uv}\theta_{-}-T_{uu}\theta_{+}\right)\,, \\
\partial_v M&=\frac{1}{n}V^\kappa_ne^fr^{n+1}\left( T_{uv}\theta_{+}-T_{vv}\theta_{-}\right)\,,
\end{align}
\end{subequations}
where $\theta_\pm$ describe the expansion rate for the null directions 
\begin{align}
\theta_{+}&=n\frac{\partial_v r}{r}\,, \qquad 
\theta_{-}=n\frac{\partial_u r}{r} \,. 
\end{align}
The variation formula (\ref{variation_uv}) does not involve the Lovelock coupling coefficients nor the 
information on the Weyl tensor of the Einstein space explicitly. This fact is advantageous for discussing the monotonicity property of the quasilocal mass as described below. 

We fix the spacetime orientation by declaring that the future-directed null vector $\partial/\partial v$ 
(resp. $\partial/\partial u$) is outgoing (reps. ingoing). Namely, $\theta_+>0$ and $\theta_-<0$ hold on an untrapped surface. Remark that each value of $\theta_\pm$ is not an invariant quantity by virtue of the remaining freedom of rescaling $u\to U(u), v\to V(v)$. Instead, $e^f \theta_+\theta_-$ enjoys an invariant physical meaning characterizing the trapping nature.

In order to extract the physically reasonable results, we impose the energy conditions on the matter fields. 
The null energy condition for the matter field implies 
\begin{align}
T_{uu}\geq 0 ,\ \ \ T_{vv}\geq 0\,, 
\end{align}
while the 
dominant energy condition for the matter field implies 
\begin{align}
T_{uu}\geq 0 ,\ \ \ T_{vv}\geq 0,\ \ \ T_{uv}\geq 0\,.
\end{align}
Note that only the information of radial directions is encoded on these inequalities.

Let us now establish that our quasilocal mass exhibits a monotonicity property, which 
 is desirable for $M$ as a physically reasonable mass function. 

\begin{Prop}[Monotonicity]
If the dominant energy condition holds, the quasilocal mass is nondecreasing along outgoing null or spacelike directions on an untrapped surface. 
\end{Prop}
The proof follows immediately from (\ref{variation_uv}).\qed

Let us next move on the the positivity claim. To this end, let us first define the regular center. The central point is said to be regular center if 
\begin{align}
\kappa -\left( Dr \right)^2 \sim Cr^2
\end{align} 
holds where $C$ is a nonvanishing constant.

\begin{Prop}[Positivity-I]
If the dominant energy condition holds and the spacetime has a regular center which is surrounded by untrapped surfaces, quasilocal mass is non-negative.
\end{Prop}
Suppose that $W(k)$ is nonvanishing. 
Around the regular center, quasilocal mass behaves as
\begin{align}
M\simeq V^{\kappa}_n \frac{1}{2^{k+1}}\frac{a_k}{k}\frac{r^{n-2k+1}}{(n-2k+1)}W(k)\,.
\label{M_center}
\end{align}
This implies 
\begin{align}
& \partial_v M\simeq V^{\kappa}_n \frac{1}{2^{k+1}}\frac{a_k}{k}\frac{r^{n-2k+1}\theta_{+}}{n}W(k) \,,
\label{Mv_center}\\
& \partial_u M\simeq V^{\kappa}_n \frac{1}{2^{k+1}}\frac{a_k}{k}\frac{r^{n-2k+1}\theta_{-}}{n}W(k)\,.
\end{align}
Combining with the monotonicity property and (\ref{Mv_center}), 
the dominant energy condition requires 
\begin{align}
a_kW(k)>0 \,.
\end{align}
This proves the positivity of quasilocal mass around the regular center.
The monotonicity property establishes the claim, as we desired.\qed

It is worth noting that for the $\kappa =1$ case, the regular center is always 
surrounded by untrapped surfaces, while this is not the case for $\kappa =-1$. 
Remark also that the Misner-Sharp mass behaves $M\propto r^{n+1}$ around the regular center for the case with $\ma K^n$ being the maximally symmetric space, whereas $M\propto r^{n-2k+1}$
for the present case.

Next, let us consider the case in which the spatial hypersurface admits a marginal 
surface as its inner boundary. On the marginal surface we have $(Dr)^2=0$; hence, the following version of the positivity holds.

\begin{Prop}[Positivity-II]
Suppose the dominant energy condition and $\Lambda \leq 0$. If the spacelike hypersurface admits a marginal surface as its inner boundary,  then the quasilocal mass admits a positive lower bound, provided that the Lovelock coefficients and Weyl tensor satisfy the following conditions for all $m$, 
\begin{align}
&a_m\left[ \sum_{l=0}^{m-1}\frac{2^{l+1}}{l+1}\Bi{m-1}{l}\left( \prod_{p=0}^{2l}(n-(2m-2)+p) \right)\kappa^{l+1}W(m-1-l)+\frac{W(m)}{m(n-2m+1)}\right] \geq 0\,.
\label{aWcond}
\end{align}
\end{Prop}
This is clear from the monotonicity and the definition of quasilocal mass.\qed

We obtained a condition (\ref{aWcond}) under which the positivity of the mass holds. 
Inspired by string theory, we may physically fix some of the Lovelock coefficients. 
However, it appears that the sign of Weyl tensor of the Einstein space is not controllable. 
It would be better if we have a clearer physical and mathematical meaning of (\ref{aWcond}). 
We leave this for future investigations.

To conclude this section, let us make a brief comment on the asymptotic behavior of the quasilocal mass. 
If the Einstein space is a round sphere, the metric falls into the standard definition of asymptotic flatness, and it would be meaningful if the asymptotic value of the quasilocal mass converges to the ADM mass, as argued in \cite{Maeda:2007uu,Maeda:2011ii}. If the Einstein space is not the maximally symmetric space, the metric exhibits a slow falloff and it does not allow asymptotically flat/AdS solutions in the standard sense. For this reason, we shall not attempt to discuss the asymptotics for the quasilocal mass.

\section{Trapping horizons}
\label{sec:trapping}

The concept of event horizon is not of practical use because the identification of its locus  requires the knowledge of the evolution of Einstein's equations into the entire future. A more convenient manner to characterize locally the strong gravity is the trapping horizon, which was originally proposed by Hayward~\cite{hayward1994}. In this section, we address some properties of trapping horizons in the present settings. 

The {\it trapping horizons} are the $n+1$-dimensional hypersurface foliated by $n$-dimensional marginal surfaces on which $\theta_+\theta_-=0$ is satisfied. Set $\theta_+=0$ on the marginal surface in what follows. Then the marginal surface is said to be {\it future} for $\theta_-<0$, {\it past} for $\theta_->0$, 
{\it outer} for $\partial_u \theta_+<0$ and {\it inner} for $\partial_u \theta_+>0$. 
By definition, the notion of trapping horizons is quasilocal and does not make any references to the asymptotic structure. One may deduce intuitively that the future-outer trapping horizons are of the most relevance for a local description of dynamical black holes, since inside the trapping horizon both of the outgoing and ingoing rays are converging. In the following discussion, we shall be mainly interested in (future-)outer trapping horizons. According to the proposition 12.2.4 of \cite{Wald}, trapped regions cannot be causally connected to null infinity, provided the null convergence condition and the cosmic censorship are valid. Therefore, the existence of the trapped regions implies the event horizon in a physically reasonable condition.

The properties of trapping horizons have been analyzed in detail for the Gauss-Bonnet gravity \cite{Nozawa:2007vq} and for the Lovelock gravity \cite{Maeda:2011ii} with the maximally symmetric horizons. In order for the trapping horizons to inherit properties in general relativity, we have to assume a certain inequality involving the Lovelock coefficients and the Weyl tensor for the Einstein space. 
 
 The next proposition specifies the causal character of the trapping horizon.
 The proof is the same as in Ref. \cite{Nozawa:2007vq}.

\begin{Prop}[Signature law]
\label{prop:sign}
Under the null energy condition the outer trapping horizon is nontimelike, provided that  the following condition on Weyl tensor and the Lovelock coefficients holds for all $m$,
\begin{align}
&a_m\left[ \sum_{l=0}^{m-1}2^{l+1}\Bi{m-1}{l}\left( \prod_{p=0}^{2l}(n-(2m-2)+p) \right)\kappa^{l}W(m-1-l)\right] \geq 0\,.
\label{cond_sig}
\end{align}
\end{Prop}
Let $\xi=\xi^v\partial_v+\xi ^u \partial_u$ be a generator of the outer trapping horizon at 
which $\theta_+=0 $ and $\partial_u \theta_+<0$.  
Since the trapping horizon is foliated by marginal surfaces, we have
\begin{align}
\label{theta_lie}
\ma L_\xi \theta_+=\xi^v\partial_v \theta_++\xi^u \partial_u \theta_+=0\,.
\end{align} 
Evaluating the ($v,v$) component of (\ref{Loveeq_tf}) at the trapping horizon we get 
\begin{align}
\label{}
T_{vv}=-\sum_{m=1}^k\sum_{l=0}^{m-1}\frac{a_m 2^{l-m+1}}{nr^{2m-2}}\Bi{m-1}{l}
 \left(\prod_{p=0}^{2l}(n-2m+2+p)\right)\kappa ^l W(m-1-l) \partial_v\theta_+ \,.  
\end{align}
The null energy condition and the inequality (\ref{cond_sig}) thus assures 
$\partial_v\theta_+ <0$. Hence (\ref{theta_lie}) implies that $\xi^u\xi^v\le 0$ is satisfied. 
If the trapping horizon is timelike, this inequality is not satisfied. We therefore arrive at the claim.\qed

The most interesting property of the event horizon of a black hole is the area increasing theorem 
(Proposition  of 12.2.6 of \cite{Wald}). It turns out that a similar property holds for the trapping horizon. 

\begin{Prop}[Area law]
\label{prop:area}
Under the null energy condition and the conditions in Proposition \ref{prop:sign}, the area of outer trapping horizon, $A(r)=V^{\kappa}_{n}r^n$, increases along its generator.
\end{Prop}
The proof directly follows from 
\begin{align}
\mathcal{L}_{\xi}A&=nr^{n-1}V^{\kappa}_{n} \left( \xi^u\partial_ur+\xi^v\partial_vr\right) \notag \\
                            &=r^{n}V^{\kappa}_n \theta_{-}\xi^u>0\,,
\end{align}
where we have used $\xi^v>0$ for the nonspacelike (spacelike) trapping horizon 
to be future-pointing (outgoing), hence $\xi^u\le 0$ from the signature law. 
This completes the proof.\qed

\subsection{Dynamics of trapping horizon}

In general relativity, the trapping horizons display laws analogous to ordinary black-hole  thermodynamics even in a dynamical circumstance~\cite{hayward1994}.  
Since the unified first law (\ref{U1st}) represents the energy balance, it can be used to deduce the 
thermodynamic first law for a trapping horizon. One can recast  (\ref{U1st})  into
\begin{align}
A\psi_a
            =&D_aM\notag \\
            &+\frac{V^\kappa_n}{2}r^nD_ar\sum_{m=1}^k  \sum_{l =0}^{m-1}\frac{a_m2^{l-m+1}}{r^{2m-2}}\left( \begin{matrix} m-1 \\ l \end{matrix}\right) \left( \frac{D^2r}{r}-{(n-2m+1)}\frac{(\kappa -(Dr)^2)}{(l +1)r^2}\right)\notag \\
            &\qquad\qquad\qquad\qquad\times\left(\prod_{p=0}^{2l }(n-(2m-2)+p)\right)(\kappa -(Dr)^2)^{l}W(m-1-l) \notag \\
                           &-\frac{V^\kappa_n}{2}r^nD_ar\sum_{m=1}^k\frac{1}{2^{m}}\frac{a_m}{m}\frac{W(m)}{r^{2m}}+V^\kappa_nr^nD_ar\Lambda
\end{align}
Let $\xi^a $ be a generator of the trapping horizon.  Hence,  along the trapping horizon 
$(Dr)^2=0$, we get 
\begin{align}
A\psi_a\xi^a=\kappa_{TH}V^{\kappa}_n\xi^aD_a\Bigg[ \sum_{m=1}^k  \sum_{l =0}^{m-1}a_m2^{l-m+1}\frac{r^{n-2m+2}}{n-2m+2}\left( \begin{matrix} m-1 \\ l \end{matrix}\right) \prod_{p=0}^{2l }(n-(2m-2)+p)\kappa^{l}W(m-1-l) \Bigg] \,, 
\label{Apsi}
\end{align}
where we have defined 
\begin{align}
\kappa_{\rm TH}\equiv \left.\frac{1}{2}D^2r \right|_{r_h} \,. 
\label{kappa_TH}
\end{align}
One can interpret (\ref{kappa_TH}) as a surface gravity of a trapping horizon, since it fulfills~\cite{hayward1998}
\begin{align}
\label{}
K^a D_{[a }K_{b]} =\kappa_{\rm TH} K_b \,,
\end{align} 
where the equality is evaluated on the trapping horizon. 
Note that this equation resembles the equation defining the surface gravity of a Killing horizon~\cite{Wald}. 
It deserves to emphasize that the surface gravity is not constant over the trapping horizon, 
as can be inferred from the Vaidya-type radiating solution (see Appendix \ref{app:matter}.). 

The unified first law reads $\delta_\xi M=Ai_\xi \psi+P\delta_\xi V$, hence 
$Ai_\xi \psi$ term should be identified as $T\delta_\xi S$ term. 
Assuming that the temperature is related to $\kappa_{\rm TH} $ by $T=\kappa_{\rm TH}/(2\pi)$, 
we can identify the entropy of a trapping horizon as
\begin{align}
S=2\pi V^\kappa_n\Bigg[ \sum_{m=1}^k  \sum_{l =0}^{m-1}a_m2^{l-m+1}\frac{r_h^{n-2m+2}}{n-2m+2}\left( \begin{matrix} m-1 \\ l \end{matrix}\right) \left(\prod_{p=0}^{2l }(n-(2m-2)+p)\right) \kappa^{l}W(m-1-l) \Bigg] \,.
\end{align}
Eq. (\ref{Apsi}) also justifies the physical interpretation of $\psi^a$ as flux current, since the change of trapping horizon entropy is responsible for the flux through the horizon. 
In the general relativistic case, the entropy is proportional to the area of the trapping horizon. 
The Lovelock black holes therefore admit a correction arising from higher-curvature terms~\cite{Myers:1988ze}. 
The highest term $l=m-1$ is also present for the maximally symmetric horizons, while the other terms represent the contribution coming from the Einstein horizon.  

Now the entropy of a trapping horizon is obtained, we move to prove the entropy increasing law. 
This corresponds to the second law of black hole dynamics. 

\begin{Prop}[Entropy law]
Under the null energy condition and the conditions in Proposition \ref{prop:sign}, the entropy of outer trapping horizon increases along its generator.
\end{Prop}
The variation of the entropy along the generator gives
\begin{align}
\mathcal{L}_{\xi}S&= 2\pi V^\kappa_n\Bigg[ \sum_{m=1}^k  \sum_{l =0}^{m-1}\frac{a_m}{n}2^{l-m+1}r_h^{n-2m+2}\left( \begin{matrix} m-1 \\ l \end{matrix}\right) \left(\prod_{p=0}^{2l }(n-(2m-2)+p)\right) \kappa^{l}W(m-1-l) \Bigg] \theta_{-}\xi^u\,.
\end{align}
The proof follows immediately from the same argument as the area theorem.\qed

\subsection{Wald's entropy}

In the previous subsection, we derived the entropy of a trapping horizon by requiring the first law of thermodynamics for the trapping horizon. Here we reproduce it by Wald's prescription for the Killing horizons~\cite{Wald:1993nt,iyerwald1994}. 

Suppose that the metric admits a nondegenerate, bifurcate Killing horizon $r=r_h$ with a bifurcation surface $B$. 
\begin{align}
\label{}
S_W&=-2\pi\int \left( \frac{\partial \mathcal{L}}{\partial R_{\mu \nu \rho \lambda}}\right) \epsilon_{\mu \nu}\epsilon_{\rho \lambda}\D V^\kappa_n \,, 
\end{align}
where $\epsilon_{\mu\nu} $ is the binormal to $B$ given by (\ref{Kodama}), 
\begin{align}
S_W        &=-2\pi\int  \sum_{m=1}^{k}\frac{1}{2^m}\frac{a_m}{m}\left( \frac{\partial\delta^{\mu_1 \mu_2 \dots \mu_{2m-1}\mu_{2m}}_{{\nu_1 \nu_2 \dots \nu_{2m-1}\nu_{2m}}}R_{\mu_1 \mu_2}{}^{\nu_1 \nu_2}\dots R_{\mu_{2m-1}\mu_{2m}}{}^{\nu_{2m-1}\nu_{2\nu}}}{\partial R_{\mu \nu \rho \lambda}}\right) \epsilon_{\mu \nu}\epsilon_{\rho \lambda}dV^\kappa_n\notag \\
        &=-2\pi\int  \sum_{m=1}^{k}\frac{1}{2^m}a_m\left( \delta^{ac\mu_3 \mu_4 \dots \mu_{2m-1}\mu_{2m}}_{{bd\nu_3 \nu_4 \dots \nu_{2m-1}\nu_{2m}}}R_{\mu_3 \mu_4}{}^{\nu_3 \nu_4}\dots R_{\mu_{2m-1}\mu_{2m}}{}^{\nu_{2m-1}\nu_{2\nu}}\right) \epsilon_{a c}\epsilon^{b d}dV^\kappa_n\notag \\
        &=2\pi V^\kappa_n\Bigg[ \sum_{m=1}^k  \sum_{\ell =0}^{m-1}a_m2^{\ell-m+1}\frac{r_h^{n-2m+2}}{n-2m+2}\left( \begin{matrix} m-1 \\ \ell \end{matrix}\right) \prod_{p=0}^{2\ell }(n-(2m-2)+p)\kappa^{\ell}W(m-1-\ell) \Bigg] .
\end{align}
Therefore, we can see that the entropy we defined from the quasilocal mass coincides with Wald's entropy.

We have derived the expressions of entropy for the Killing horizons and see that it coincides with 
the stationary limit of trapping horizons. One can alternatively utilize the Kodama vector instead of the  generator of the Killing horizon to directly derive for the trapping horizon as demonstrated in \cite{Hayward:1998ee}.

\section{Final remarks}
\label{sec:conclusion}

In this paper we explored various properties of the spacetimes which are the warped product of 
a two-dimensional Lorentzian spacetime and an $n$-dimensional Einstein space. Assuming the form of the stress-energy tensor to be (\ref{SETensor}), we revealed that the Weyl curvature of the Einstein space must obey certain conditions (\ref{Wcond}). This assumption comes not only from the simplification, but also from the requirement that the metric (\ref{metric}) admits a vacuum solution. Some nontrivial examples are given in Appendix~\ref{app:ex}. We found that all the isotropy irreducible spaces fulfill this property. Our study enlarges considerably the solution space of Lovelock gravity.

One immediate conclusion for replacing the $n$-dimensional maximally symmetric subspace by the Einstein space is that the metric shows the fall-off behaviors different from the standard one. This means that the $M-r_h$ diagram for the static black hole is much more complicated than \cite{Whitt:1988ax}. A possible future work in this direction is to examine the $P-V$ criticality of a black hole with Einstein horizons and to expose the thermodynamic phase structure. 

We then proceeded to define a quasilocal mass and explored its physical properties, by 
extending the previous works~\cite{Maeda:2007uu,Maeda:2011ii}. The rederivation of the quasi-local mass in terms of the Kodama flux is desirable for the physical interpretation of the quasilocal mass. 
Up to the certain conditions among the Lovelock coefficients and the Weyl curvature of the Einstein space, it turns out that the quasilocal mass shares the same behavior as that in general relativity \cite{Hayward:1994bu}. This implies that the Misner-Sharp-type quasilocal mass continues to be useful also in Lovelock gravity and can be utilized to obtain a coherent picture of spacetime dynamics as exemplified by gravitational collapse. We hope to come back to the point for the deeper mathematical and physical understanding of the conditions (\ref{aWcond}), (\ref{cond_sig}). 

Our formulation of Lovelock solutions with the warped $n$-dimensional Einstein space is very robust and has plenty of potential applications. We expect that the geometrodynamics approach to Hamiltonian formulation of Lovelock black holes~\cite{Kunstatter:2012kx} can be extended to the case with Einstein horizons. It is also interesting to consider the effects of nonlinear Maxwell field~\cite{Maeda:2008ha} and higher-rank $p$-form fields~\cite{Bardoux:2010sq}, which would display an intriguing thermodynamic phase by the interplay with Weyl curvatures of Einstein spaces. One may also explore the generalization of C functions~\cite{Anber:2008js} and the maximal entropy principle~\cite{Cao:2014fka} into the present context.

\section*{Acknowledgements}

We are grateful to Hideki Maeda and Marcello Ortaggio
for useful comments. 
S.O. was supported by a JSPS Grant-in-Aid for Scientific Research No. 25-9997.
This work was partly supported by JSPS and INFN.

\appendix 

\section{Examples of Einstein spaces}
\label{app:ex}

We give some examples of Einstein spaces satisfying (\ref{Wcond}).
The condition (\ref{Wcond2}) is trivially met for the homogeneous spaces $G/H$ since 
they admit a frame in which the curvature tensors take constant values. 
Remark that not all the homogeneous spaces fulfill (\ref{Wcond1}). 
For instance, we find that the homogeneous Sasaki-Einstein space $T^{(1,1)} $ described by the coset 
${\rm SU}(2)\times {\rm SU}(2)/{\rm U}(1)$~\cite{Klebanov:1998hh}  fails to satisfy the  $m=2 $ condition of (\ref{Wcond1}). 
It has been also shown in \cite{Dotti:2005rc} that the family of homogeneous Bohm metrics does not satisfy (\ref{Wcond2}). 
The following examples are either the products of maximally symmetric spaces or the symmetric spaces.

\subsection{Products of maximally symmetric spaces}

\subsubsection{$K^p \times K^q$}
Let us consider the space consisting of the two products of maximally symmetric spaces
$K^p\times K^q$ ($p+q=n$), where $K^p$ denotes the $p$-dimensional maximally symmetric space 
with a sectional curvature $\kappa=0, \pm 1$. Then the $n$-dimensional Einstein metric reads
\begin{align}
\label{}
\gamma_{ij}\D x^i \D x^j =\frac{p-1}{p+q-1}\D \Omega^2 (K^p)
+\frac{q-1}{p+q-1}\D \Omega^2 (K^q)\,.
\end{align}
Decomposing indices $i,j,...$ into $\{A,B,...\}$ for $\Omega^p$ and $ \{I,J,...\}$ for $\Omega^q$, 
the nonvanishing components of the Weyl tensor are 
\begin{align}
\label{}
C_{AB}{}^{CD}=\frac{\kappa q}{p-1} \delta_{AB}^{CD} \,, \qquad 
C_{AI}{}^{BJ}=-\kappa \delta_A^B\delta_I^J \,, \qquad 
C_{IJ}{}^{KL}=\frac{\kappa p}{q-1} \delta_{AB}^{CD} \,. 
\end{align}
In order to see whether the constraints (\ref{Wcond1}) are satisfied, it is enough to check that $W(m)^i_j$ is proportional to $\delta^i_j$ for all $m$.
The constraint for $m=1$ is trivial, whereas the $m=2$ imposes the condition obtained in \cite{Dotti:2005rc} and requires $p=q$. For arbitrary order, the left-hand side of (\ref{Wcond1}) reads
\begin{align}
\label{}
W(m)^i_j=&\kappa^m \delta^i_j\frac{p}{p-1} \sum_{s=0}^m\sum_{t=0}^{m-s}
2^{2m-(s+t)}\Bi{m}{s}\Bi{m-s}{t}\bigg( \frac{p}{p-1}\bigg)^{s+t}
\frac{(-1)^{m-(s+t)}}{[p-(m+s-t)-1][p-(l-s+t)]}
\notag \\
&\quad \times 
\left(\prod_{\alpha=0}^{m+s-t}(p-(m+s-t)-1+\alpha)\right)
\left(\prod_{\beta=0}^{m+t-s}(p-(m-s+t)+\beta)\right).
\label{Wij_SpxSq}
\end{align}
This proves that the constraints (\ref{Wcond1}) are satisfied at arbitrary order of $m$.

\subsubsection{$K^p\times \cdots \times K^p$}

The $q$th products of maximally symmetric spaces $K^p$ also satisfy the condition (\ref{Wcond}). 
The metric in this space is given by
\begin{align}
\label{}
\gamma_{ij}\D x^i \D x^j=\sum_{\sigma=1}^q \left(\frac{p-1}{pq-1}\right)\D \Omega^2(K^p_{(\sigma)})\,. 
\end{align}
The Weyl tensor is computed to give
\begin{align}
\label{}
C_{i^{(\sigma)}j^{(\sigma)}}{}^{k^{(\sigma)}l^{(\sigma)}}=\kappa \frac{p(q-1)}{p-1} 
\delta^{k^{(\sigma)}l^{(\sigma)}}_{i^{(\sigma)}j^{(\sigma)}}\,, \qquad 
C_{i^{(\sigma)}j^{(\rho)}}{}^{k^{(\sigma)}l^{(\rho)}}
=-\kappa \delta_{i^{(\sigma)}}^{k^{(\sigma)}}\delta_{j^{(\rho)}}^{l^{(\rho)}} \quad \textrm{for $\sigma \ne \rho$},
\end{align}
where $\{ i^{(\sigma)}, j^{(\sigma)}, \dots \}$ are indices for $\Omega^2(K^p_{(\sigma)})$. After lengthy calculations, the left-hand side of (\ref{Wcond1}) is 
\begin{align}
\label{}
W(m)^i_j=&\kappa^m \delta_{i}^{j} \left(\frac{pq}{pq-2m}\right)
\sum_{\sum_{1\le \sigma\le \rho\le q}a_{(\sigma)(\rho)}=m}
\Biggl[ \frac{m!}{\prod_{1\le \zeta\le\eta\le q}a_{(\zeta)(\eta)}!} \left(\frac{2p(q-1)}{p-1}\right)
^{\sum^q_{\xi=1}a_{(\xi)(\xi)}} (-4)^{\sum_{1\le \pi <\tau \le q}a_{(\pi)(\tau)}}
\notag \\ 
&\quad \times \left(\frac{1}{p-(a_{(\gamma)(\gamma)}+\sum_{\beta=1}^qa_{(\beta)(\gamma)})-1}\prod_{\chi_\gamma=0}^{a_{(\gamma)(\gamma)}+\sum_{\beta=1}^qa_{(\beta)(\gamma)}}\left[p-\left(a_{(\gamma)(\gamma)}+\sum_{\beta=1}^qa_{(\beta)(\gamma)}\right)-1+\chi_\gamma\right]\right)\notag \\
&\quad \times \prod_{\alpha=1, \alpha\ne \gamma}^q\left(\frac{1}{p-(a_{(\alpha)(\alpha)}+\sum_{\beta=1}^qa_{(\alpha)(\beta)})}
\prod_{\chi_\alpha=0}^{a_{(\alpha)(\alpha)}+\sum_{\beta=1}^qa_{(\alpha)(\beta)}}\left[p-\left(a_{(\alpha)(\alpha)}+\sum_{\beta=1}^qa_{(\alpha)(\beta)}\right)+\chi_\alpha\right]\right)
\Biggr]\,. \label{Wij_Spxq}
\end{align}
where $a_{(\sigma)(\rho)}$ ($1\le \sigma, \rho \le q$) are numbers satisfying 
$\sum_{1\le \sigma \le \rho \le q}a_{(\sigma)(\rho)}=m$.  One can easily see that the right-hand side of (\ref{Wij_Spxq}) is independent of any choice of $\gamma$. This proves that the constraints (\ref{Wcond1}) are satisfied at arbitrary order of $m$.

\subsection{Isotropy irreducible spaces}

We discuss here a class of homogeneous spaces satisfying (\ref{Wcond}). Let us start by introducing some mathematical definitions (see~\cite{Besse} for details). Let $G $ be the semisimple Lie group. A manifold is said to be homogeneous when $G$ acts transitively. Letting $H$ denote the isotropy subgroup of $G$, the homogeneous space is described by a coset space $G/H$. The homogeneous space is said to be symmetric if there exists an involutive automorphism $\sigma$ such that $\sigma(h)=h$, $\sigma(k)=-k$, where $h$ is the Lie algebra of $H$ and $k$ is its complement in the Lie algebra $g$ of $G$; i.e. we have a Cartan decomposition, $g=h\oplus k$. The homogeneous space is isotropy irreducible if the linear isotropy representation of $H$ is irreducible and the symmetric isotropy irreducible space is simply called an irreducible symmetric space. 

As we commented earlier, the first condition (\ref{Wcond1}) is always satisfied for any homogeneous spaces $G/H$, whereas not all the homogeneous spaces meet the second condition (\ref{Wcond2}). 
Let us here remind the theorem that any symmetric $G$-invariant rank-two tensors are unique up to the multiplication constant in the isotropy irreducible spaces~\cite{MuellerHoissen:1987cq}. It follows that the symmetric tensor $W(m)_{ij}$, as well as the Ricci tensor,  constructed out of the $G$-invariant metric must be proportional to the $G$-invariant metric in the isotropy irreducible spaces. Therefore, we have proven that all the isotropy irreducible $G$-homogeneous spaces are Einstein and satisfy the required conditions for the Weyl tensor (\ref{Wcond}). 
In what follows, we shall provide some examples  by focusing on the Riemannian irreducible symmetric spaces\footnote{As far as the noncompact case is concerned, this exhausts all the possibilities (Proposition 7.46 of \cite{Besse}).} for which the explicit metrics are known. Of course, all the $G$-homogeneous spaces listed in Tables 7.102--7.107 of Besse's textbook \cite{Besse} provide desired examples.

\subsubsection{Complex projective space $\mathbb{CP}^N$}
\label{sec:CPN}

The complex projective space is a K\"ahler-Einstein 
symmetric space $\mathbb{CP}^N={\rm SU}(N+1)/{\rm SU}(N)$. 
In the complex coordinates, the K\"ahler potential is given by
\begin{align}
\label{}
K=\ell^2 \ln \left(1+\sum_{I=1}^N  z^I \bar z^I \right) \,, 
\end{align}
The standard Fubini-Study metric 
$\D s^2=2 \gamma_{I\bar J}\D z^I \D \bar z^J$ with 
$\gamma_{I\bar J}=\partial^2 K/\partial z^I\partial \bar z^J$ reads
\begin{align}
\label{}
\D s^2= \frac{2\ell^2}{1+|z|^2} 
\left(\D z^I \D \bar z^I -\frac{|\bar z^I \D \bar z^I|^2}{1+|z|^2}\right) \,,
\end{align}
where $I=1,...,N=n/2$. The curvature tensors in the complex basis are given by
\begin{align}
\label{CPn_curvatures}
R^I{}_J{}^K{}_L[\gamma]=-2\ell^{-2}\delta^I{}_{(J}\delta^K{}_{L)} \,, \qquad 
R^I{}_J[\gamma]=(N+1) \ell^{-2} \delta^I{}_J\,, \qquad 
R[\gamma]=2N(N+1)\ell^{-2}\,. 
\end{align}
Hence, $\ell^2=(n+2)/[2(n-1)]$ gives the unit normalization. 
In the real basis, we have
\begin{align}
\label{}
C_{ij}{}^{kl}[\gamma]=\ell^{-2}\left(-\frac{3}{n-1}\delta_{[i}^k\delta_{j]}^l
+J_{ij}J^{kl}+J_i{}^{[k}J_j{}^{l]}\right) \,, 
\end{align}
where $J$ is the complex structure satisfying $J_i{}^kJ_k{}^j=-\delta_i{}^j$ and 
$J_{ij}=J_{[ij]}$. 
Hence, it is almost obvious to see that the symmetric tensor 
$W(m)_{ij}=W(m)_{ji}$ is proportional to $\gamma_{ij}$ for each $m$, although we omitted the detailed computations. 

One can also consider the Bergmann space, which is the negative curvature version of $\mathbb{CP}^N$; 
namely, it is described by a coset ${\rm SU}(N-1,1)/{\rm S}({\rm U}(N-1)\times {\rm U}(1))$. 
This metric arises as nonlinear sigma models of vector and hyper multiplets in the framework of $\ma N=2$ supergravity~\cite{FVP}. 
The K\"ahler potential is given by
\begin{align}
\label{}
K=-\ell^2 \ln \left(1-\sum_{I=1}^N  z^I \bar z^I \right) \,, 
\end{align}
Since the curvature tensors of the Bergmann space are given by 
the Wick rotation $\ell \to i \ell$ of (\ref{CPn_curvatures}), 
it is obvious that the Bergmann space also satisfies (\ref{Wcond}), as the complex projective space.

\subsubsection{Quadratic surface $Q^{N-2}(\mathbb C)={\rm SO}(N+2)/{\rm SO}(N)\times {\rm SO}(2)$}

The quadratic surface is the compact space of positive curvature in $n=2N$ dimensions. 
The noncompact version ${\rm SO}(N,2)/{\rm SO}(N)\times {\rm SO}(2)$ appears as the nonlinear sigma model for $\ma N=2$ supergravity~\cite{Ferrara:2007pc}.  
The K\"ahler potential is given by
\begin{align}
\label{}
K=\ell^2 \log \left(
1+2\sum_I ^Nz^I\bar z^I+\sum_{I,J}^N(z^I\bar z^J)^2 
\right) \,. 
\end{align}
The curvature tensors in the complex basis are
\begin{align}
\label{}
R^I{}_J{}^K{}_L=-\ell^{-2}(\delta^I{}_J\delta^K{}_L+\delta^I{}_L\delta^K{}_J-\delta^{IK} \delta_{JL}) \,, 
\end{align}
giving $\ell^2=n/[4(n-1)]$. 
It is then straightforward to show explicitly that the condition (\ref{Wcond}) holds.

\subsubsection{$G_{2(2)}/{\rm SO}(4)$}

This is a negative curvature, noncompact space and describes a 
nonlinear sigma model arising from 
 the ${\rm U}(1)^2$ reduction of five-dimensional minimal supergravity. 
Setting $\xi\to \ln f$ and $f\to -f$ [this sign flip ensures $H={\rm SO}(4)$ instead of ${\rm O}(2,2)$] in Ref. \cite{Bouchareb:2007ax}, 
the metric is given by 
$\D s^2=\ell^2 \sum_{i=1}^8(e^i )^2$, where 
\begin{align}
\label{}
e^1=&\frac{\D f}{\sqrt 2 f}\,, \qquad e^2=\frac{1}{\sqrt 2f}(\D \chi +v_1\D u_1+v_2 \D u_2)\,, 
\qquad e^3=\sqrt{\frac 32}\D \phi \,, \qquad e^4=\sqrt{\frac 32}e^\phi \D \kappa \,, \notag \\
e^5=& \frac 1{\sqrt{2f}}e^{-3\phi/2}\D v_1 \,, \qquad 
e^6=\frac 1{\sqrt{2f}}e^{3\phi/2}[\D u_1+\kappa^3\D v_1-\sqrt 3\kappa (\D u_2-\kappa\D v_2)]\,, 
\\
e^7=& \frac 1{\sqrt{2f}}e^{-\phi/2}(\D v_2+\sqrt 3\kappa\D v_1) \,, \qquad 
e^8=\frac 1{\sqrt{2f}}e^{\phi/2}(\D u_2-2\kappa \D v_2-\sqrt 3\kappa^2 \D v_1)\,.\notag 
\end{align}
Here eight variables ($f, \chi, \phi, \kappa, v_1,u_1, v_2, u_2$) span the coordinate basis. 
The Ricci tensor is $R_{ij}[\gamma] =-4 \ell^{-2}\gamma_{ij}$; hence, 
$\ell^2=4/7$ gives the correct normalization. From the general argument given at the beginning of this appendix, this space satisfies (\ref{Wcond}), which can also be checked explicitly (the quartic term is trivial due to the dimensionally dependent identities~\cite{Lovelock2}).  One can also confirm that the curvature tensors take constant values in the above orthogonal basis.

\subsubsection{${\rm SL}(N, \mathbb R)/{\rm SO}(N)$}

This is the negative curvature symmetric space of dimension $n=\frac 12(N+2)(N-1)$. 
The coset arises as a nonlinear sigma model when one dimensionally reduces the $D$-dimensional Einstein's gravity on an $N$-dimensional torus~\cite{Maison:1979kx}. The metric reads
\begin{align}
\label{SLnR}
\gamma_{ij}\D x^i \D x^j =-\frac {\ell^2}4 {\rm Tr}(\D M \D M^{-1}) \,, 
\end{align}
where $M$ is an $N\times N$ symmetric matrix with ${\rm det}M=1$. 
The Ricci tensor is given by $R_{ij}[\gamma] =-(N/\ell^2) \gamma_{ij}$, hence 
$\ell^2=2N/(N^2+N-4)$.  
A convenient parametrization of the matrix $M$ is
$M=U_R^TU_DU_R$ where $U_D$ is the diagonal unimodular matrix and $U_R$ is the upper triangle matrix with the diagonal entries equal to unity. To the best of our knowledge, the Riemann curvature of this  space does not seem to have a simple expression for general $N$. In spite of this, the group-theoretical reason ensures (\ref{Wcond}), as can be checked for small $N$ explicitly.

\section{Inclusion of matter fields}
\label{app:matter}

We present in this appendix some solutions with matter sources. 
These solutions are instructive to give some physical insights for readers.

\subsection{Maxwell field}

Let $F_{\mu\nu}$ be a Faraday two-form satisfying 
\begin{align}
\label{Feq}
\D F=0 \,, \qquad \D \star F=0 \,, 
\end{align}
with a stress-energy tensor 
\begin{align}
\label{}
T_{\mu\nu}=2 \left(F_{\mu\rho}F_\nu{}^\rho -\frac 14 g_{\mu\nu}
F_{\rho\sigma}F^{\rho\sigma}\right) \,.
\end{align}
Due to the restriction $T_{ai}=0$, the permissible form of the two-form reads 
\begin{align}
\label{}
 F = \frac 12 E \epsilon_{ab} \D y^a \we \D y^b +\frac 12\hat F_{ij} \D x^i \we \D x^j \,.
\end{align}
From the field equations, we get $E=E(y)$, $\hat F_{ij}=\hat F_{ij}(x^k)$ with 
\begin{align}
\label{}
E=\frac{Q_e}{r^n}\,,\qquad \D \hat F=0 \,, 
\qquad \D \star _\gamma \hat F=0 \,,
\end{align}
where $Q_e$ is a constant. 
Hence, $\hat F$ is a harmonic two-form on $\ma K^n$. The nontrivial 
$\hat F$ exists provided the second Betti number of $\ma K^n$ is nonvanishing. 
Moreover, the stress-tensor of the electromagnetic field
must fall into the form (\ref{SETensor}), which yields the additional restrictions 
\begin{align}
\label{F_cond}
\hat F_{ik}\hat F_j{}^k \propto \gamma_{ij} \,, \qquad 
\hat F_{ij}\hat F^{ij}={\rm const. } 
\end{align}
The existence of nontrivial $\hat F$ satisfying these conditions also puts a restriction on the Einstein manifold.  For instance, 
the condition (\ref{F_cond}) is satisfied if $\ma K^n$ is even dimensional and admits an almost complex structure $J$ with $\hat F_{ij}\propto  J_{ij}$. 
Note that in odd dimensions we have ${\rm det}\hat F_{ij}\equiv 0$; hence, the first condition of (\ref{F_cond}) implies $\hat F_{ij}=0$ provided $\ma K^n$ is Euclidean (this can be shown by taking the determinant of 
$\gamma^{kl}\hat F_{ik}\hat F_{jl}=n^{-1}\gamma_{ij}\hat F_{kl}\hat F^{kl}$. See e.g. \cite{Ortaggio:2007hs} for a similar analysis). 

Denoting $\hat F_{ik}\hat F_{j}{}^k=Q_m^2 \gamma_{ij}$ where $Q_m$ is a constant,
we have
\begin{align}
\label{}
T_{ab}=-\left(\frac{Q_e^2}{r^{2n}}+\frac{nQ_m^2}{2r^4}\right)g_{ab} 
\,, \qquad 
T_{ij}=\left(\frac{Q_e^2}{r^{2n-2}}+\frac{(4-n)Q_m^2}{2r^2}\right)\gamma_{ij} \,.
\end{align}
Since $\hat T_{ab}=0$ is fulfilled, it follows that Birkhoff's theorem holds for the present field configurations. The solution of $(Dr)^2 >0$ is given by (\ref{staticsol}) with
the left-hand side of (\ref{Mass_static}) replaced by 
\begin{align}
\label{}
\mu \to \mu  +V_n^\kappa \left(\frac{Q_e^2}{(1-n)r^{n-1}} +\frac{nQ_m^2}{2(n-3) r^{3-n}}\right) \,.
\end{align}
Remark that we have $Q_m=0$ for odd $n$.
One sees that the nonvanishing $Q_m$ gives rise to the slow falloff of the metric 
at infinity. 

\subsection{Vaidya solution}

Another interesting solution of physical interest  is the Vaidya-type 
radiating solution~\cite{Kobayashi:2005ch,Maeda:2005ci,Nozawa:2005uy,Cai:2008mh}. 
Let us consider the null dust fluid as the stress-energy tensor 
\begin{align}
\label{}
T_{\mu\nu}=\rho l_\mu l_\nu \,, 
\end{align}
where $l_\mu$ is a null vector and $\rho $ is the energy density of the null dust. Here, 
let us introduce the outgoing null coordinates:
\begin{align}
\label{}
\D s_2^2=-f(v,r)\D v^2 +2 \D v \D r \,, \qquad l_\mu =-\nabla_\mu v \,.
\end{align}
It follows that $P=0$, $\hat T_{ab}=\rho \nabla_a v \nabla_b v $; 
hence, the unified first law implies $\D M=\rho A \D v$. This means that $M$
is a function of $v$ only and the metric function $f(v,r)$ satisfies
\begin{align}
\label{Mass_Vaidya}
M (v)= &V_n^\kappa \Biggl[\sum_{m=1}^k \frac{1}{2^{m+1}}\frac{a_m}m \frac{r^{n-2m+1}}{n-2m+1}
W(m)-\frac{r^{n+1}}{n+1}\Lambda \nonumber \\
& +\sum_{m=1}^k \sum_{l=0}^{m-1}2^{l-m}a_m \frac{r^{n-2m+1}}{l+1}\Bi{m-1}{l}
\left(\prod_{p=0}^{2l}(n-(2m-2)+p)\right) (\kappa-f)^{l+1}W(m-1-l)\Biggl]\,.
\end{align}
$\rho (v,r)$ is given by
\begin{align}
\label{}
\rho= \frac 1{V_n^\kappa r^n}\frac{\partial M(v)}{\partial v} \,. 
\end{align}
The null energy condition requires that $M(v)$ is an increasing function of $v$.



\end{document}